\documentclass[11pt]{article}

\usepackage{graphicx}
\usepackage{ifthen}
\usepackage{float}
\usepackage{amsmath}
\usepackage{amssymb}
\usepackage{amscd}
\usepackage{amsfonts}
\usepackage{graphicx}
\usepackage{epsfig}
\usepackage{fullpage}
\usepackage{subfigure}
\usepackage{citesort}
\usepackage{amsthm}

\def \labs{{|}}

\def\reals{\hbox{I\kern -.19em R}}

\def\realf{\hbox{I\kern -.19em F}}
\def\reale{\hbox{I\kern -.19em E}}
\def\proof{\noindent {\bf Proof.} }

\def \o{\over}
\def \al{\alpha}

\def \la{\lambda}

\def \eps{\epsilon}

\def \prob #1 {{\large {\bf {#1}}} \hspace{5mm}}
\def \part #1 {{\bf {#1}} \hspace{3mm}}

\newfont{\bbb}{msbm10}

\newcommand{\argmin}{\mbox{argmin}}

\newcommand{\card}{\mbox{Card}}

\newcommand{\ca}{\mbox{${\cal A}$}}

\newcommand{\ci}{\mbox{${\cal I}$}}

\newtheorem{theorem}{Theorem}
\newtheorem{lemma}{Lemma}
\newtheorem{definition}{Definition}

\newtheorem{proposition}{Proposition}

\newtheorem{corollary}{Corollary}
\newtheorem{example}{Example}

\newtheorem{problem}{Problem}

\newenvironment{dfn}{\begin{definition}
\begin{em}}{\end{em}\end{definition}}
\newenvironment{thm}{\begin{theorem}
\begin{em}}{\end{em}\end{theorem}}
\newenvironment{lmm}{\begin{lemma}
\begin{em}}{\end{em}\end{lemma}}

\newenvironment{prp}{\begin{proposition}
\begin{em}}{\end{em}\end{proposition}}

\newcommand{\remark}{{\bf Remark: \hspace{0.1in}}}

\newcommand{\bdm}{\begin{displaymath}}
\newcommand{\edm}{\end{displaymath}}
\newcommand{\beq}{\begin{equation}}
\newcommand{\eeq}{\end{equation}}
\newcommand{\bqry}{\begin{eqnarray}}
\newcommand{\eqry}{\end{eqnarray}}
\newcommand{\bqryo}{\begin{eqnarray*}}
\newcommand{\eqryo}{\end{eqnarray*}}
\newcommand{\bary}{\begin{array}}
\newcommand{\eary}{\end{array}}

\newcommand{\g}{\gamma}

\newcommand{\nom}{\nonumber}

\newcommand{\proofover}{\hfill\vrule height8pt width6pt depth 0pt\newline}

\def\5sp{\; \; \; \; \; }
\begin{document}
\date{}
\title{Deterministic Designs with Deterministic Guarantees: Toeplitz Compressed Sensing Matrices, Sequence Design and System Identification \thanks{This research was supported by ONR Young Investigator Award N00014-02-100362 and a Presidential Early Career Award (PECASE), NSF
CAREER award ECS 0449194, and NSF Grant CCF 0430983 and
CNS-0435353}
\author{Venkatesh Saligrama\\
Department of Electrical and Computer Engineering\\
Boston University, Boston MA 02215\\
E-mail: srv@bu.edu.}}
\maketitle

\begin{abstract}
In this paper we present a new family of discrete sequences having ``random like'' uniformly decaying auto-correlation properties. The new class of infinite length sequences are higher order chirps constructed using irrational numbers. Exploiting results from the theory of continued fractions and diophantine approximations, we show that the class of sequences so formed has the property that the worst-case auto-correlation coefficients for every finite length sequence decays at a polynomial rate. These sequences display doppler immunity as well. We also show that Toeplitz matrices formed from such sequences satisfy restricted-isometry-property (RIP), a concept that has played a central role recently in Compressed Sensing applications. Compressed sensing has conventionally dealt with sensing matrices with arbitrary components. Nevertheless, such arbitrary sensing matrices are not appropriate for linear system identification and one must employ Toeplitz structured sensing matrices. Linear system identification plays a central role in a wide variety of applications such as channel estimation for multipath wireless systems as well as control system applications. Toeplitz matrices are also desirable on account of their filtering structure, which allows for fast implementation together with reduced storage requirements.
\end{abstract}
\section{Introduction} \label{s.intro}

Sequences with special properties are required in a number of applications ranging from communication systems, radar systems and system identification. Many of these applications require families of sequences with low auto-correlation, cross-correlation and resiliency to doppler offsets. In communication systems signature sequences with low auto-correlation properties have been employed in wireless commmunications (see \cite{proakis}) to overcome self interference due to multi-path effects and co-channel interference due to multi-access communications respectively.

Toeplitz structured matrices naturally arise in linear system identification, a problem that is at the core of many applications ranging from channel estimation in multipath wireless systems to model estimation in control applications. In wireless systems the channel coefficients are sparse and channel estimation is identical to compressed sensing with Toeplitz structured matrices. In parallel in several control applications, signals with low auto-correlation sequences are required  when a model parametrization is generally unknown and models of increasing complexity are adaptively chosen~\cite{risanen} to characterize the system. This usually requires a persistently exciting input of arbitrary order. In addition, an input probing signal that is ``optimal'' to any model order is preferable and this leads us to seek aperiodic input sequences with good non-circular autocorrelation properties.  It can also arise in the context of system identification in the presence of unmodeled dynamics and noise. There is a wide array of literature dealing with system identification in the presence of unmodeled dynamics (see for example \cite{milan96} and references therein). The solution to this problem requires designing inputs that can suppress the corruption due to both noise and unmodeled errors~\cite{srvsep}. While noise can be overcome by applying a persistent input, to suppress the effect of unmodeled dynamics requires aperiodic input sequences whose non-circular (or zero-padded) autocorrelation uniformly decreases to zero. It turns out~\cite{srvsep,srvconv} that the contribution from the unmodeled dynamics in the parametric error is directly related to the rate of decay of the worst-case non-circular autocorrelation coefficients.

In many applications such as multipath wireless systems~\cite{bajwa:ciss08} as well as other mediums such as acoustic/RF the system can be modeled as a finite impulse response (FIR) sequence. It turns out that in these cases the support of FIR~\cite{bajwa:ciss08} is sparse. Consequently, this leads to the problem of reconstruction of sparse FIR systems.

Sparse reconstruction has been extensively dealt with in the context of Compressed Sensing (CS)~\cite{candes:tit06a,candes:tit06b,candes:tit05,donoho:tit06}. CS involves recovering a sparse signal $x$ from linear observations of the form $y = Ux$. Suppose the true signal, $x$, has fewer than $k$ non-zero components the solution to the $\ell_0$ problem recovers this solution if and only if every sub-matrix of $U$ formed by choosing $2k$ arbitrary columns of $U$ has full column rank. Unfortunately, it turns out that the general solution is intractable. In \cite{candes:tit05,candes:tit06b,candes:tit06a,donoho:tit06,tropp:tit04} it is shown that for sufficiently small $k$, the $\ell_1$ problem is equivalent to solving the $\ell_0$ problem whenever the sensing matrix $U$ satisfies the so called RIP property. The RIP property of sparsity $k$ amounts to the requirement that the singular values of every sub-matrix formed by selecting $k$ columns of $U$ is close to one. The question now arises as to what matrices satisfy RIP property. Earlier work on CS established RIP property for random constructions and recent work has produced deterministic designs based on deterministic sequence constructions~\cite{devore,calderb}.

Nevertheless, this setup is not directly applicable to the sparse FIR system identification problem.  Specifically, in this setting the output is the convolution of inputs with the channel coefficients. This amounts to taking a product of Toeplitz matrix of the inputs and a vector consisting of FIR coefficients as components. Therefore, this calls for RIP property for Toeplitz matrices. Motivated by these issues~\cite{aeronssp07} consider random Toeplitz matrices and derive information theoretic limits while \cite{bajwa:ciss08,bajwa:ssp07} develop RIP properties for random generated Toeplitz matrices. In this paper we design deterministic Toeplitz matrices having guaranteed RIP properties.

To account for compressed sensing in the context of system identification we first present a new family of discrete sequences having ``random like'' uniformly decaying auto-correlation properties. The new class of infinite length sequences are higher order chirps constructed using irrational numbers. Exploiting results from the theory of continued fractions and diophantine approximations, we show that the class of sequences so formed has the property that the worst-case auto-correlation coefficients for every finite length sequence decays at a polynomial rate. We also show that Toeplitz matrices formed from such sequences satisfy restricted-isometry-property (RIP). Linear system identification plays a central role in a wide variety of applications such as channel estimation for multipath wireless systems as well as control system applications. Toeplitz matrices are also desirable on account of their filtering structure, which allows for fast implementation together with reduced storage requirements. Additionally, we show that these sequences are immune to doppler offsets.

The organization of the paper is as follows. In Section~\ref{sec:prob} we motivate Toeplitz structured matrices in CS and System Identification. We then formulate the problem of CS in terms of seeking sequences with autocorrelation properties. This motivates the design of sequences which is dealt with in Section~\ref{sec:seqdesign}. The following Section~\ref{sec:mrip} deals with matrix constructions that further improve upon the RIP properties. We finally conclude with a brief discussion of practical and implementational issues.

\section{Problem Setup} \label{sec:prob}

Linear system identification problems are generally characterized by measured output, $y$, the input, $u$, Gaussian noise, $w$, and a linear time-invariant system, $G$ that are related by the following discrete time equation.
\beq \label{eq:expt}
y(t)=Gu(t)+w(t),\,\,\,t=0,\,1,\,\ldots,\,n,\,\,
\eeq
where, the system $G$ is identified by its kernel, $\{g_t\}$, which is usually assumed to be an element of $\ell_1$, where, $\ell_1$ is identified with the space of bounded analytic functions on the unit disc~\cite{srvsep}. The objective is to estimate $G$ in some metric norm. Obviously, with finite data arbitrary elements of $\ell_1$ cannot be estimated. One typically assumes in this context that the system can be decomposed into a finite dimensional model, $H$ and a small residual unmodeled dynamics, $\Delta$. If the finite parametrization happens to be the class of finite impulse response sequences (FIR) of order $m$ we obtain in expanded notation,
\beq
y_s=Gu_s+w_s=\overbrace{\sum_{k=0}^m g_k u_{s-k}}^{\mbox{FIR Model}} +
\overbrace{\sum_{k=m+1}^s g_k u_{s-k}}^{\mbox{Residual Error}} + w_s
\eeq
The task is to estimate $H \equiv \{g_0,\,g_1,\,\ldots,\,g_m\}$ given that $\Delta \equiv (g_{m+1},g_{m+2},\ldots.)$ has $\ell_1$ norm smaller than $\gamma$.

Further expanding into matrix notation we get,
\beq \label{eqn:triangletoep}
\left ( \begin{array}{c} y_1 \\ y_2 \\ \vdots \\ y_n \end{array} \right )=
\left [ \begin{array}{cccc}
u_0&0&\ldots&\ldots\\u_1&u_0&0&\ldots\\\vdots&\ddots&\ddots&\vdots\\u_{n-1}&u_{n-2}&\ldots&u_0
\end{array} \right ] \left ( \begin{array}{c} g_1 \\ g_2 \\ \vdots \\ g_n \end{array} \right ) + \left ( \begin{array}{c} w_1 \\ w_2 \\ \vdots \\ w_n \end{array} \right )
\eeq
Now in the absence of noise it is clear that a {\it pulse input} of unit-amplitude is sufficient
to recover the FIR model exactly. On the other hand persistent noise can only be averaged out by a persistent input. However, a persistent input will also "excite" the unmodeled error. In particular for the above equation, whenever, $s \geq m+1$, the data also contains contributions from the unmodeled dynamics. For a {\it periodic input}, $u$, with $u_{s+l}=u_s$, the data
for $s= 0,\,l,\,2l,\,\ldots$ can be written as:
$$
y_s=\sum_{k=0}^s g_k u_{s-k} + w_s = \left ( g_0 + g_1+\ldots \right
) u_0 + \left (g_1 + g_{l+1}+\ldots\right ) u_{1}+ \ldots + w_s
$$
Thus one obtains information only on linear combinations of the
$g_k$'s and not the individual coefficients. It is therefore impossible to determine only the model-coefficients no matter how long the input signal and how large the length of the period. Therefore, no matter how large the period, in the worst case the unmodeled error will couple with the model-set dynamics.
Although we describe the simple case of FIR models and residual error, similar decompositions can be obtained for more general parameterizations~\cite{srvsep,srvconv}.

We define aperiodic autocorrelations function to characterize the design problem for robust identification.
\begin{dfn}
A n-window aperiodic autocorrelation function (ACF) of a real/complex valued sequence, $\{u_t\}_{t \in Z^+}$, is defined as:
$$
\tilde r_u^n(\tau)={1\o n}\sum_{s=0}^{n-\tau} u_s u^*_{s+\tau},\,\tau \in Z
$$
where the asterisk denotes the complex conjugate.
\end{dfn}
It turns out that a sufficient condition (see \cite{srvconv} for necessary condition) on input sequences for estimation of the optimal finitely parameterized model is that the worst-case aperiodic autocorrelation asymptotically approach zero.
$$
\max_{0<\tau\leq n} {|\tilde r_u^n(\tau)| \over |\tilde r_u^n(0)|} \longrightarrow 0
$$

The special case when the unmodeled dynamics, $\Delta$, is negligible is of importance in many applications.
Here the impulse response terminates after $p$ coefficients and an input of length $n$ is usually applied. Consequently, we get
$$
y_s=\sum_{k=0}^s g_k u_{s-k} + w_s,\,\, s \leq p
$$
and
$$
y_s=\sum_{k=0}^p g_k u_{s-k} + w_s,\,\, s > p
$$
This corresponds to a ``fat'' Toeplitz matrix of the form:
\begin{align}
\label{eqn:fulltoep}
    \hspace{-5ex} U =
        \begin{bmatrix}
            u_0\\
            u_1         & \ddots\\
            \vdots      & \ddots    &   & u_0\\
                        &           &   & u_1\\
            u_{n-1}         &           &   & \vdots\\
                        & \ddots\\
                        &           &   & u_{n-1}
        \end{bmatrix}.
\end{align}
where the number of columns in the above matrix is $p$.

There is an other situation that arises in the FIR context when the output response is observed in steady state. Here, the input $u_t$ is assumed to start from time $t= -\infty$ and the output is observed between time $t=p$ until $t=p+n-1$. In this situation we get,
$$
y_s=\sum_{k=0}^p g_k u_{s-k} + w_s
$$
Organizing the above equation as a matrix for $s \geq p$ leads to the following Toeplitz matrix:
\begin{align}
\label{eqn:toepmat}
        U \ = \
        \begin{bmatrix}
            u_p         & u_{p-1}   & \hdots    & u_2       & u_1\\
            u_{p+1}     & u_{p}     & \hdots    & u_3       & u_2\\
            \vdots      & \vdots    & \ddots    & \ddots    & \vdots\\
            u_{p+n-1}   & u_{n+k-2} & \hdotsfor{2}          & u_n
        \end{bmatrix},
\end{align}
Note that the structure of the steady state matrix leads to time-dependent auto-correlations, which we define below:
\begin{dfn}
An n-window autocorrelation function (ACF) of a real/complex valued sequence, $\{u_t\}_{t \in Z^+}$, is defined as:
$$
r_u^n(t,\tau)={1\o n}\sum_{s=t}^{t+n} u_s u^*_{s+\tau},\,\, \tau \in Z
$$
where the asterisk denotes the complex conjugate.
\end{dfn}
So far the discussion above has focused on different aspects of system identification. We will now relate this to compressed sensing problems. These problems arise when the support of FIR coefficients is sparse.
As we described in the previous section a central ingredient typically employed in establishing that convex programming algorithms lead to sparse recovery is the so called restricted isometry property (RIP). We describe the RIP property next. We follow closely the development in \cite{aeronita08}.

Let $D$ be the diagonal matrix consisting of $\ell_2$ norms of each column of $U$. Let, $\ci_q$ be the collection of all subsets of $\{1,\,2,\,\ldots,\,p\}$ of cardinality $q$. Enumerate the elements of this sub-collection as $l=1,\,2,\,\ldots, {n \choose q}$ and let $\pi(l)$ denote the lth element of $\ci_q$.  $U_{\pi(l)},\,D_{\pi(l)}$ denote the sub-matrices  of $U,\,D$ respectively obtained by selecting columns with indices in $\pi(l)$.

Let, $\Sigma_l$ be the normalized correlation matrix, i.e.,
$$
\Sigma_l = D^{-1/2}_{\pi(l)} U_{\pi(l)}^T U_{\pi(l)} D^{-1/2}_{\pi(l)}
$$

%

We next define the RIP property.
\begin{dfn}[RIP property] \label{dfn.rip}
A $n \times p$ sensing matrix, $U$, characterized by the triple $(n,p,q)$ is said to have RIP property of order $q$ if there is a number $\delta_q \in [0,1)$ such that every correlation matrix $\Sigma_l$ satisfies:
$$(1-\delta_q)\leq \lambda_{\min}(\Sigma_l) \leq \lambda_{\max}(\Sigma_l) \leq (1+\delta_q), \,\, \forall \,\, l = 1,\,2,\,\ldots, {n \choose q}$$
\end{dfn}

Note that RIP property of order $q$ immediately implies RIP property of smaller order. This follows from the observation~\cite{aeronita08} that for $\tilde \pi(l) \subset \pi(l)$ we get
$$
\lambda_{\min}(U_{\pi(l)}^T U_{\pi(l)}) \leq \lambda_{\min}(U_{\tilde \pi(l)}^T U_{\tilde \pi(l)}),\,\,\lambda_{\max}(U_{\tilde \pi(l)}^T U_{\tilde \pi(l)}) \leq \lambda_{\max}(U_{\pi(l)}^T U_{\pi(l)})
$$

We are now left to establish eigenvalue bounds for the correlation matrices of all orders.
First note that the coefficients of the correlation matrix are the autocorrelation coefficients. As remarked earlier the autocorrelation coefficients are different for the different matrices. In particular for Equation~\ref{eqn:toepmat} we have,
$$
\Sigma_l = \left [{r_u^n(t,s-t) \over r_u^n(t,0)} \right ]_{t,s};\,\, t,\,s \in \pi(l) \in \ci_q
$$
On the other hand for Equation~\ref{eqn:fulltoep} we get
$$
\Sigma_l = \left [ {\tilde r_u^n(t-s) \over \tilde r_u^n(0) } \right ]_{t,s};\,\,t,\,s \in \pi(l) \in \ci_q
$$

We can now state a sufficient condition for RIP property based on eigenvalue bounds for correlation matrices of \cite{haykin}. These bounds are based on a straightforward application of Gersgorin theorem. The reader is also referred to Theorem~2 of \cite{bajwa:ciss08} for a closely related result derived for random Toeplitz matrix constructions.
\begin{thm} \label{thm:rip}
Suppose $U$ is Toeplitz matrix of Equation~\ref{eqn:toepmat}.
Then $U$ satisfies the RIP property of order $q$ if
$$
R = \max_{\pi(l) \in \ci_q} \sum_{t, s \in \pi(l),\, s \not = t} {|r_u^n(t,s-t)| \over |r_u^n(t,0)|}< 1
$$
Now suppose $U$ is a Toeplitz matrix as in Equation~\ref{eqn:fulltoep}. Then $U$ satisfies the RIP property of order $q$ if
$$
R = \max_{\ci_q} \sum_{\tau \in \ci_q} {|\tilde r_u^n(\tau)| \over |\tilde r_u^n(0)|}< 1
$$
\end{thm}
\proof
The proof follows by a straightforward application of Gershgorin's theorem applied to Correlation matrices (see Page 60 in \cite{haykin}). In particular from \cite{haykin} it follows that the eigenvalues of the correlation matrix $\Sigma_l$ can be upper and lower bounded by the sum of the off-diagonals. In other words, if
$$
R_l = \sum_{t, s \in \pi(l),\, s \not = t} {|r_u^n(t,s-t)| \over |r_u^n(t,0)|}
$$
then,
$$
\lambda_{\min}(\Sigma_l) \geq 1 - R_l,\,\,  \lambda_{\max}(\Sigma_l) \leq 1 + R_l
$$
Now to ensure every correlation matrix satisfies the RIP property we take the maximum over all the elements in $\ci_q$. The condition for Equation~\ref{eqn:fulltoep} follows in an identical fashion and is omitted. \proofover


Our first attempt at establishing RIP property will be by quantifying the worst-case autocorrelation coefficient. In particular in Section~\ref{sec:seqdesign} we will describe sequences whose worst-case autocorrelation coefficients decay at a polynomial rate. In the following Section~\ref{sec:mrip} we will directly deal with matrix constructions.
\begin{dfn} \label{def:regularacf}
A bounded sequence, $\{u(t)\},\,t=1,\,2,\,\ldots,\,n+p-1$, is said to have a uniformly polynomial decay (PDACF) property if there is a $\g >0$ such that the worst-case autocorrelation satisfies:
$$
\left | {r_u^n(t,\tau) \o r_u^n(t,0)} \right | \leq C n^{-\g},\,\, \forall \, t=1,\,2,\,\ldots,\,p,\,\,0 \leq t+\tau \leq p
$$
\end{dfn}
\begin{corollary}
The Toeplitz matrix $A$ of Equation~\ref{eqn:toepmat} RIP property of order $q = {\cal O}(n^{\g})$ if the sequence satisfies PDACF property with decay coefficient $\g$.
\end{corollary}
Note that the PDACF property may not hold for arbitrary number of columns $n$ and we make this precise later.

\begin{figure}
\begin{center}
\resizebox{!}{3in}{\includegraphics{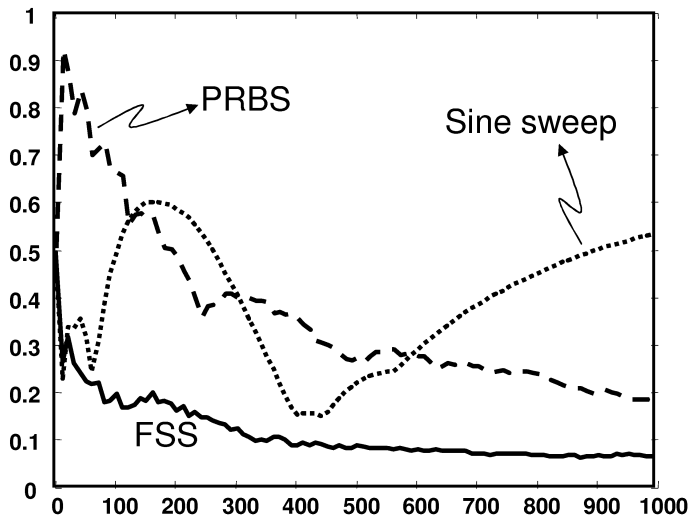}}\\
\caption{Comparison of Worst-case auto-correlation functions for various lengths of a $2^{15}-1$ order PRBS, sine-sweeps and Higher-order-chirps or fast sine sweeps (FSS).}
\label{fi:prbs}
\end{center}
\end{figure}

In summary a simple means to realize RIP property is to obtain characterizations with low autocorrelation. It can be argued that for all practical purposes one could possibly take truncations of pseudo-random binary sequences (PRBS) sequences with arbitrary long period. However, it turns out that truncations of PRBS sequences do not necessarily have small autocorrelations. Unfortunately the autocorrelation expressions in Theorem~\ref{thm:rip} involves correlations of truncated sequences. In Figure~\ref{fi:prbs} the worst-case auto-correlations of truncated PRBS of period $2^{15}-1$ has been plotted. As seen such truncated sequences can have poor autocorrelations. Finally, it is perhaps surprising that the widely used sine-sweep,
$u(t)=\exp(i\alpha t^2),\,\,\, \al \in \reals, \,\,\,\, t=0,\,1,\,2,\ldots$, does not meet the requirements either. In \cite{srvsep}, we show that,
$$
\limsup \max_{0<\tau\leq n} |\tilde r_u^n (\tau )|\geq 1/2
$$
The basic reason is that the auto-correlation function turns out to be equal to $\sin(n\tau \al/2)/(n \sin(\tau \al))$ and subsequences $\tau_j$, and $n_j$ can be suitably chosen so that the limiting value of $\tau_j \al \mod(2\pi)$ tends to zero

We next provide sufficient conditions in terms of worst-case decay of autocorrelation coefficients.
\begin{dfn} \label{def:aperiodic}
A bounded sequence, $\{u(t)\},\,t=1,\,2,\,\ldots,\,n$, is said to have an aperiodic uniformly polynomial decay (PDACF) property if there is a $\g >0$ such that the worst-case aperiodic autocorrelation satisfies:
$$
\left | {\tilde r_u^n(\tau) \o \tilde r_u^n(0)} \right | \leq C n^{-\g},\,\, \forall \, \tau=1,\,2,\,\ldots,\,n
$$
\end{dfn}
\begin{corollary}
The Toeplitz matrix $A$ of Equation~\ref{eqn:fulltoep} has an RIP property of order $q = {\cal O}(n^{\g})$ if the sequence satisfies PDACF property with decay coefficient $\g$.
\end{corollary}
Note that the PDACF property may not hold for arbitrary number of columns $n$ and we make this precise later.

\section{Sequence Design} \label{sec:seqdesign}
It turns out that higher-order-chirps (HOC) sequences are PDACF sequences. These are deterministic sequences of the form $$u(t)=\exp(i2\pi\al t^3)$$ where $\al$ is some irrational number. In applications the real part of the complex sequence will actually be applied and to simplify our exposition we consider the complex valued signal here.
\begin{thm} \label{thm:c6chirp3}
The third order chirp sequences has the PDACF property for the ACF function as defined in Definition~\ref{def:regularacf} with $\gamma = 0.25$ for $\tau \leq \lambda n$ for all quadratic irrational numbers.
\end{thm}
\noindent
\remark Recall that algebraic numbers are those numbers that are roots of any polynomial defined over a field of integers.

\noindent
\remark Note that the higher-order chirps are complex valued. However, identical properties hold for real-part and imaginary part of these signals.

The proof is broken down into several steps. In the first step we will upper bound the ACF function by means of simpler functions.

Denote by $\|x\|$ the distance of $x$ from the closest integer:
\beq \label{eq:c6dirr}
\|x\| = \min_{A \in Z} |x-A|
\eeq
and
$$\lbrack x \rbrack = \argmin_{ A \in Z} |x-A|$$

The following elementary facts follows:
\begin{prp} \label{prp:props}
The following hold:
\begin{eqnarray*}
\|kx\|&=&\|k\|x\|\|\\
\labs \sin(2\pi x)\labs &\in& \left [{\|x\|\o 2}, \|x\| \right ]
\end{eqnarray*}
\end{prp}
\proof
To prove the first assertion we note that:
$$
\lbrack x+k\rbrack = \lbrack x \rbrack + k,\,\, \,\,\forall\,\,k\in Z^+
$$

Next we proceed as follows:
\begin{eqnarray*}
\|kx\|&=&\labs kx-\lbrack kx \rbrack \labs \\&=& \labs kx-\lbrack kx - k \lbrack x \rbrack + k\lbrack x \rbrack \rbrack \labs = \labs kx-\lbrack k(x - \lbrack x \rbrack) \rbrack - k\lbrack x \rbrack\labs  \\&=& \labs k(x - \lbrack x \rbrack)-\lbrack k(x - \lbrack x \rbrack) \rbrack \labs
\end{eqnarray*}
To obtain bounds for the sinusoid function, we observe that:
$$|\sin(2\pi x)| = |\sin(2\pi\|x\|)| \in [0.5\|x\|,\|x\|]$$
where the last assertion is a standard result in elementary calculus.
\proofover

With these preliminaries we can reduce the expression for ACF in terms of the function $\|\cdot\|$ as in the following lemma.
\begin{lmm}
The ACF function, $|r_u^n(\tau)|,\,\,\tau \in Z^+$, for the HOC satisfies the following inequality for any irrational number $\al \in \reals$.
$$
\left | r_u^n(t,\tau) \right |^2 \leq {1 \o n}\sum_{k=0}^p {3 \o 1 + n \|k \tau \alpha\| },\,\, 1 \leq \tau \leq p = \lambda n
$$
and for the aperiodic ACF
\beq \label{e.acfbdd}
\left | \tilde r_u^n(\tau) \right |^2 \leq {3 \o n}\sum_{k=0}^n {1 \o 1 + n \|k \tau \alpha\| },\,\, 1 \leq \tau \leq n
\eeq
\end{lmm}
\proof First, the ACF function and its aperiodic counterpart can be simplified by straightforward algebraic manipulations as follows:
$$
\left | \tilde r_u^n(\tau) \right |^2 \leq {1\o n^2}
\sum_{k=1}^{n-\tau} \left | {\sin(2\pi \al k \tau (n-\tau)) \o
\sin(2\pi \al k \tau)} \right |  + {C \o n} = {C \o n} + {1\o n^2}
\sum_{k=1}^L \left | {\sin(2\pi L k x) \o \sin(2\pi k x)} \right |,\,\, L=n-\tau,\,\,x=\tau\al, 1 \leq k,\tau \leq n
$$
Similarly,
$$
\left | r_u^n(t,\tau) \right |^2 \leq {1\o n^2}
\sum_{k=1}^{n} \left | {\sin(2\pi \al k \tau n) \o
\sin(2\pi \al k \tau)} \right |  + {C \o n} = {C \o n} + {1\o n^2}
\sum_{k=1}^n \left | {\sin(2\pi n k x) \o \sin(2\pi k x)} \right |,\,\,x=\tau\al,\,\,1 \leq k,\tau \leq p = \lambda n
$$

To further simplify these expressions we consider the function, $H_L(\cdot)$:
$$
H_L(x)={3 L \o 1 + L \|x\|}
$$
and obtain the following property.
\begin{prp} \label{prp:nfunc}
The function $H_L(x)$ is a monotonic function over $L$, i.e., $H_j(x) \leq H_k(x)$ for $j<k$. Furthermore, it satisfies:
$$
\left | {\sin(2\pi L x) \o \sin(2\pi x)} \right | \leq H_L(x)
$$
\end{prp}
\proof
Consider the function, $H_L(\cdot)$:
\beq
H_L(x) = {3L \o 1+L\|x\|}
\eeq
Since, $|\sin(2\pi Lx)/\sin(2\pi x)|\leq L$, it follows
$$
\left |{(1+L\|x\|) \sin(2\pi Lx)\o \sin(2\pi x)} \right | \leq
L + L {\|x\| \o |\sin(2\pi x)|} |\sin(2\pi L x)| \leq 3L
$$
where for last inequality we have used Proposition~\ref{prp:props}  \proofover \\
Next, we show that $H_L(\cdot)$ is monotonic in $L$. Suppose, we have given two integers, $n,\,m$, with $n<m$, it follows that:
$$
m+mn\|x\| \leq n+mn\|x\| \Longleftrightarrow {m\o 1+m\|x\|} \leq {n\o 1 + n\|x\|}
$$
The result now follows by inspection. \proofover

Note that the upper bounds for both cases are identical for $\lambda=1$, i.e., $p=n$. For simplicity we consider $\lambda =1$ case and comment on what happens when $\lambda > 1$ later. The main difficulty now in establishing the result is that $\liminf_q \|q \al\| =0$ for every real number, i.e., $\|j\al\|$ comes close to zero infinitely often. The proof therefore rests on the fact that very few terms in the sequence $(\|\tau\al\|,\,\|2\tau\al\|,\,\ldots,\,\|n\tau\al\|$ are close to zero for any phase $0<\tau\leq n$. A well known result in continued fraction expansion theory~\cite{khinchin} provides how closely can rational numbers approximate irrational numbers:
\begin{thm} \label{thm:khinchin}
If $\alpha$ is a quadratic irrational number(i.e. irrational solution of a quadratic polynomial) there exists a constant $C_\alpha > 0$ such that,
$$
{C_\al \o j} \leq  \|j\al\|,\,\, j \in Z^+
$$
Furthermore, for every constant $\eps>0$, for almost all irrational numbers $\al \in (0,1)$ except on set of lebesgue measure zero, the inequality,
$$
\|q \al \| < {1 \over q \log^{1+\eps} q},\,\, q \in Z^+
$$
has only a finite number of solutions.
\end{thm}
\remark To put things into perspective we point out that $C_\al = {1 \o \sqrt{5}}$ for the golden ratio, $\al = (1+
\sqrt{5})/2$.

Unfortunately, this inequality alone is insufficient for deriving the upper bounds for worst-case ACF, as shown below:
$$
\max_{0< \tau \leq n} {1\o n}\sum_{k=0}^n  {1 \o 1+n\|k\tau\al\|} \leq {1\o n}\sum_{k=1}^n  {1 \o 1+{nC_{\al} \o k n}} \not \longrightarrow 0
$$
Consequently, we need to rely on more intricate properties of continued fraction theory to develop an accurate estimate.

The above setup suggests that it is convenient to deal with the family of $n$ irrational numbers, $\tau \al$ with $0< \tau \leq n$. We denote by the symbol $\beta$ any irrational number in this family. Specifically, we let $$\beta \in {\cal S}(\alpha,n)=\{ \|\tau \alpha)\|; \,\, \tau=1,\,2,\,\ldots,n\}$$
Note that by employing Proposition~\ref{prp:props} we can further simplify the expression in Equation~\ref{e.acfbdd} as follows:
$$
{1 \o n} \sum_{k=1}^n {1 \o 1 + n \|k \tau \al\| } = {1 \o n} \sum_{k=1}^n {1 \o 1 + n \|k (\|\beta\|) \| } ,\,\, \mbox{for some}\,\, \beta \in {\cal S}(\alpha,n)
$$
Consequently, we need to show that:
$$
\left ({1\o n}\sum_{k=0}^n  {1 \o 1+n\|k\beta\|} \right )^{1/2}\leq n^{-\gamma},\,\, \forall \,\,\beta \in {\cal S}(\alpha,n)
$$
We summarize properties from elementary continued fraction theory for the purpose of completion next. These properties and the accompanying notation have been adopted from \cite{rockett}.

\begin{figure}
\begin{center}
\resizebox{!}{1in}{\includegraphics{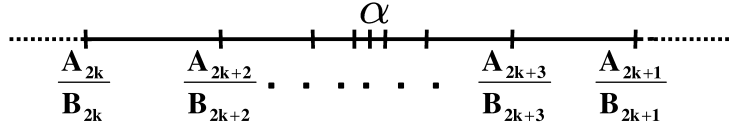}}\\
\caption{Illustration of how even and odd convergents approach an irrational number $\al$}
\label{fi:cfe}
\end{center}
\end{figure}

\paragraph{Continued Fractions:}
The continuous fraction expansion of any positive irrational number $\beta$ is given by:
$$
\beta = [a_0; a_1,\,a_2,\,\ldots,\,a_j,\,\ldots]=a_0 + {1 \o a_1 + {1 \o a_2 +  {1\o \ldots a_j \ldots}}},\,\, a_j \in Z^+
$$
where for the sake of brevity the first expression is typically used to denote the expansion. In our context we have $\beta < 1/2$ and so $a_0 = 0$.
By truncating the continuous fraction expansion we obtain a sequence of rational numbers called convergents, i.e.,
$$
{A_k \o B_k} = [a_0; a_1,\,a_2,\,\ldots,\,a_k]
$$
The numerator and denominator of the convergents satisfy the following recursion:
\bqry \label{eq:crecur}
A_{k+1}&=& a_{k+1} A_{k} + A_{k-1},\,\,\, A_{-1}=1;\,\,A_0=a_0,\,\,k \geq 0 \\
B_{k+1} &=& a_{k+1} B_{k} + B_{k-1},\,\,\, B_{-1}=0;\,\,B_0=1,\,\,k \geq 0
\eqry
Since, $a_k$s are positive integers the numerator and denominator sequences $B_k,\,A_k$s form a strictly monotonically increasing sequence. It follows that:
\beq \label{eq:gratio}
B_k = a_k B_{k-1} + B_{k-2} = (a_k a_{k-1}+1) B_{k-2} + B_{k-3} \geq 2 B_{k-2} \geq 2^{(k-1)/2}
\eeq
where the last equation follows from the fact that $a_k \geq 1$.
\paragraph{Optimal Approximation:}
The convergents are optimal rational approximants, in that, the kth convergent is the best approximation to the irrational number among all rationals which have a smaller denominator:
\beq
\|B_k\beta\| \leq \|j \beta\|,\,\, \forall \,\, 0<j \leq B_{k},\,\, k \geq 1
\eeq
For convenience we denote by,
$$
D_k = B_k \beta - A_k
$$
The $D_k$s follows a simple a recursion identical to Equations~\ref{eq:crecur}, i.e.,
\beq \label{eq:drecur}
D_k = a_kD_{k-1} + D_{k-2},\,\,D_{-1}=-1
\eeq
It follows that the so called even convergents, $A_{2k}/B_{2k}$, monotonically approach $\beta$ from the right while the odd convergents, $A_{2k+1}/B_{2k+1}$, approach $\beta$ from the left as shown in Figure~\ref{fi:cfe}.
It turns out that the value of $D_k$ is bounded from above and below by the size of the denominators given by:
\beq \label{eq:lbd}
{1\o 2B_{k+1}} \leq {1\o B_k + B_{k+1}}\leq |D_k| \leq {1\o B_k}
\eeq
\paragraph{Ostrowski Decomposition:}
In the so called Ostrowski's representation, each integer, $m$, is expressed as a linear combination of the $B_k$s. It turns out that this representation is unique in the following sense.  Any integer $m \in Z^+$ with $B_p \leq m \leq B_{p+1}$ can be uniquely decomposed as:
\beq \label{eq:ostr}
m=\sum_{j=0}^p c_{j+1} B_j
\eeq
with $0\leq c_{j+1} \leq a_{j+1}$ for $j\geq 1$ and $0 \leq c_1 < a_1$. Moreover, $c_j=0$ if $c_{j+1}=a_{k+1}$. This decomposition is obtained by first dividing $m$ with the largest denominator, $B_p$, smaller than $m$ and then decomposing the remainder with a corresponding largest denominator and so on.
\paragraph{Lower Bounds to Rational Approximation:} The significance of Ostrowski representation follows from upper and lower bounds. First note that due to translation invariance we have,
$$
\|m \beta \| = \| \sum_{j=0}^p c_{j+1} B_j \beta \| = \| \sum_{j=0}^p c_{j+1} B_j \beta - \sum_{j=0}^p c_{j+1} A_j\| =  \|\sum_{j=0}^p c_{j+1} D_j \|
$$
The following Lemma from \cite{rockett} (Lemma 1 Chap. 2) will be useful:
\begin{lemma} \label{lem.rock}
$$
|\sum_{j=0}^p c_{j+1} D_j| \geq |(c_{\gamma(m)+1} - 1)D_{\gamma(m)}| + |D_{\gamma(m)+1}|
$$
\end{lemma}

To derive bounds we need to define the type of an integer $m$. Let, $m=\sum_{j=0}^p c_{j+1} B_j$, be the Ostrowski representation for $m$. The type $\g(m)$ of an integer $m$ is an integer defined as:
$$
\g(m)\,=\,\min\{j ~|~ c_{j+1}\not = 0\}
$$
We also define $\g_n^*$ to be the maximum possible type in $Z_n = \{1,\,2,\,\ldots,\,n\}$, i.e.,
$$
\g_n^*= \max\{\g(m)~|~ m = 1,\,\ldots,\,n\}
$$
\begin{prp} \label{prp.maxtype}
The maximum type, $\g_n^*$, in the set $Z_n$ is smaller than $2\log_2 (2n)$.
\end{prp}
\proof If $m\in Z_n$ is of type $\g(m)$, then, from the ostrowski decomposition it follows that
$$
m \geq B_{\g(m)} \geq 2^{{\g(m)-1 \o 2}} \Longrightarrow 2\log_2 m +1 \geq \g(m)
$$
where, we have used Equation~\ref{eq:gratio} for the second inequality. Thus, the largest possible type in $Z_n$ is $\la_n=2\log_2 n$. Since, type is a positive integer ranging from $0$ to $2\log_2 n$, the result follows by inspection. \proofover \\

The type of the integer, $q$, signifies how small $\|q\beta\|$ can get.
We have the following result from \cite{rockett} (Theorem~1 Chap. 2) that precisely characterizes this connection:

\begin{thm} \label{thm.rock}
Let $0<\beta < 1/2$ and the integer, $m$, have type $\gamma(m)\geq 1$. Then,
$$
\|m\beta\| = |\sum_{j=0}^p c_{j+1} D_j|
$$
\end{thm}

The above results lead to the following precise characterization of how type is associated with $\|m \beta\|$:
\begin{lemma} \label{lem.lbd}
Let $0< \beta< 1/2$ be an irrational number and $m>1$ a positive integer. Then,
$$
\|m\beta\| \geq \left \{ \begin{array}{cc} \min \left ({(c_1-1)|D_0|}+{(a_2-c_2)|D_1|}, {(a_1- c_1)|D_0|} +{c_2|D_1|}\right )\\ \geq \min \left ({c_1-1\o 2 B_1}+{a_2-c_2 \o 2 B_2}, {a_1- c_1 \o 2B_1} +{c_2 \o 2 B_2}\right )& \g(m)= 0\\&\\{c_{\g(m)+1}-1\o 2 B_{\g(m)+1}}+{a_{\g(m)+2}-c_{\g(m)+2} \o 2 B_{\g(m)+2}} & \g_n^*-1 \geq \g(m)\geq 1 \\ & \\ c_{\g_n^*+1}\|B_{\g_n^*}\beta\|&\g(m)=\g_n^* \geq 1\\ & \\ \end{array} \right .
$$
\end{lemma}
\proof
The case of $\gamma(m) = \gamma_n^*$ is a direct consequence of Theorem~\ref{thm.rock}.
To establish the second case we proceed by noting that for $\gamma(m) \geq 1$ we have $\|m\beta\| = | \sum_{j=0}^p c_{j+1} D_j|$ and from the fact that the even and odd $D_k's$ have opposite sign we get:
\begin{eqnarray*}
|\sum_{j=0}^p c_{j+1} D_j| & = & |c_{\g(m)+1} D_{\g(m)}+ c_{\g(m)+2} D_{\g(m)+1}+ c_{\g(m)+3} D_{\g(m)+2}\\ &&+\ldots+c_{\g(m)+2j} D_{\g(m)+2j-1} + c_{\g(m)+2j+1} D_{\g(m)+2j}+\ldots | \\ &\geq& | c_{\g(m)+1} D_{\g(m)}+ c_{\g(m)+2} D_{\g(m)+1} + a_{\g(m)+4} D_{\g(m)+3} + a_{\g(m)+6} D_{\g(m)+5}+ \ldots |  \\&\geq& | c_{\g(m)+1} D_{\g(m)}+a_{\g(m)+2} D_{\g(m)+1}+(c_{\g(m)+2}-a_{\g(m)+2}) D_{\g(m)+1} + a_{\g(m)+4} D_{\g(m)+3}\\  && + a_{\g(m)+6} D_{\g(m)+5}+\ldots |
\end{eqnarray*}
The second inequality follows from the fact that we have removed all the terms with the same sign as $D_{\g(m)}$ (i.e. terms of the form $D_{\g(m)+2j}$) while taking the maximum possible values for terms of opposite sign. The third inequality is just a restatement of the third term. Next we utilize Equation~\ref{eq:drecur} to make the substitution that $D_k - D_{k-2} = a_k D_{k-1}$ and get,
\begin{eqnarray*}
|\sum_{j=0}^p c_{j+1} D_j| &\geq& |c_{\g(m)+1} D_{\g(m)}+(c_{\g(m)+2}-a_{\g(m)+2}) D_{\g(m)+1}+a_{\g(m)+2} D_{\g(m)+1}\\ && + a_{\g(m)+4} D_{\g(m)+3} + a_{\g(m)+6} D_{\g(m)+5}+\ldots |\\ &\geq& | c_{\g(m)+1} D_{\g(m)} + (c_{\g(m)+2}-a_{\g(m)+2}) D_{\g(m)+1} \\ &&+ (D_{\g(m)+2}-D_{\g(m)})+(D_{\g(m)+4}-D_{\g(m)+2}) + \ldots |\\ &\geq& |(c_{\g(m)+1}-1) D_{\g(m)}+(c_{\g(m)+2}-a_{\g(m)+2}) D_{\g(m)+1}|
\end{eqnarray*}
Finally, we know from Equation~\ref{eq:lbd} that
$$
|D_{\g(m)}| \geq {1\o B_{\g(m)}+B_{\g(m)+1}} \geq {1\o 2B_{\g(m)+1}}
$$
Next observing that $D_{\g(m)}$ and $D_{\g(m)+1}$ are of opposite signs the result follows.

Finally, to establish the first case, i.e., $\gamma(m)=0$ we note that a lower bound for $|\sum_{j=0}^p c_{j+1} D_j|$ follows identically as the case for $\gamma(m) \geq 1$.
We next derive an upper bound by excluding all terms of opposite sign (except for the term containing $D_1$) as $D_0$ and include the maximum possible terms of the same sign. Through algebraic manipulations as for the lower bound and noting that the even $D_k's$ are of the same sign we get:
$$
|\sum_{j=1}^{\g_n^*} c_{j+1} D_j | \leq | c_1 D_0 + c_2 D_1 + a_3 D_2 + a_5 D_4 + \ldots |  \leq | c_1 D_0 + (c_2-1) D_1|
$$
Next, substituting Equation~\ref{eq:drecur} for $D_1 = a_1 D_0 -1$ we get:
$$
|\sum_{j=1}^{\g_n^*} c_{j+1} D_j| \leq |1+(c_1-a_1) D_0+c_2 D_1|
$$
Now it follows that the latter term is strictly less than $1$ by noting that $D_0 = \beta > 0$ and $D_1$ is negative and:
$$|1+(c_1-a_1) D_0+c_2 D_1| \leq |1 + ((a_1-1)-a_1)D_0| \leq |1-D_0| \leq 1$$

We also note that $D_2=a_2D_1+D_0>0$, and so, $1+(c_1-a_1) D_0+c_2 D_1 = c_1 D_0 + (c_2 - 1) D_1 \geq c_1 D_0 + (a_2 - 1) D_1 \geq (c_1 -1)D_0 -D_1+ D_2 > 0$. Thus we have a upper and lower bound for $|\sum_{j=1}^{\g_n^*} c_{j+1} D_j|$ that are both positive and smaller than $1$, i.e.,
$$
0<{(c_1-1) D_0} + {(c_2-a_2)D_1}\leq |\sum_{j=1}^{\g_n^*} c_{j+1} D_j| \leq 1+(c_1-a_1) D_0+c_2 D_1 < 1
$$
Now noting that $D_0$ is positive and $D_1$ is negative establishes the first inequality for $\g(m)=0$. Also substituting the lower bounds of Equation~\ref{eq:lbd}
it follows that the second inequality for $\g(m)=0$ is also satisfied.
\proofover

To complete the above set of results we need a bound for the situation when $\gamma(m)=\gamma_n^* =0$. We have the following corollary:
\begin{corollary} \label{cor.lbd}
Let $0< \beta< 1/2$ be an irrational number and $m>1$ a positive integer and $\gamma(m)=\gamma_n^* =0$. Then,
$$
\|m\beta\| \geq \min \left ( (c_1-1)|D_0|, (a_1-c_1)|D_0| \right )
$$
\end{corollary}

We are now ready to prove the main theorem by a combination of the above well-known properties in continued fraction theory. The main outline of the proof is as follows. Lemma~\ref{lem.lbd} and Corollary~\ref{cor.lbd} points to the fact that the type of a number controls the value of $\|m\al\|$. This motivates partitioning of the set $Z_n=\{1,2,\ldots,\, j,\,\ldots,\, n\}$ based on its type and computing the contributions for each type. This leads us to defining the following sets:
\bqry \nom
\ca_{l,c}&= &\{ m \in Z_n ~|~ \g(m)=l,\,\,c_{\g(m)+1}=c,\,\,\mbox{in the Ostrowski expansion for m}\} \\ \nom
\ca_{l,c,d}&= &\{ m \in Z_n ~|~ \g(m)=l,\,\,c_{\g(m)+1}=c, c_{\g(m)+2}=d\,\,\mbox{in the Ostrowski expansion for m}\}
\eqry

We show in the sequel that the cardinality decreases exponentially with the type.
\begin{lmm}\label{lmm:card}
The cardinality of $\ca_{l,c}$ in the set $Z_n$ is given by:
$$
\#\ca_{l,c} \leq \left \{ \begin{array}{cc} 2\left \lfloor{n \o B_{l+1}}\right \rfloor & l < \g_n^* \\ 1 & l=\g_n^* \end{array} \right .
$$
and the cardinality of the set, $\ca_{l,c_1,c_2}$ is bounded by the cardinality of the set, $\ca_{l+1,c_2}$.
\end{lmm}
\proof
The proof requires the following proposition.
\begin{prp} \label{prp:order}
Given, two positive integers, $q,\,r$ and their corresponding Ostrowski representations, $\{c_j(q)\},\,\{c_j(r)\}$, it follows that:
$$
q < r \Longleftrightarrow \exists \, l \in Z^+ \,\,\mbox{such that}\,\,c_l(q) < c_l(r);\,\, \,\, c_j(q) \leq c_j(r),\,\, \forall \, j \geq l
$$
\end{prp}
\proof
($\Longleftarrow$ case) Suppose there is an $l$ satisfying the hypothesis. It then follows that,
\bqryo
r-q&=&\sum_{j=0}^p c_{j+1}(r) B_j - \sum_{j=0}^p c_{j+1}(q) B_j\\&=&\sum_{j=0}^{l-2} (c_{j+1}(r)-c_{j+1}(q)) B_j + \sum_{j=l-1}^p (c_{j+1}(r)-c_{j+1}(q)) B_j\\&>& -\sum_{j=0}^{l-2} c_{j+1}(q) B_j + B_{l-1}
\eqryo
It remains to show that the last term is positive. We do this by induction. Clearly, the hypothesis is true for $l=1$. Suppose, the induction hypothesis is true for $l=k$, then for $l=k+1$, we have:
$$
B_{k+1} = a_{k+1} B_k + B_{k-1} \geq \left \{ \begin{array}{c} (a_{k+1}-1) B_k + \sum_{j=1}^{k-1} c_{j+1} B_j,\,\, c_k \not = 0 \\ a_{k+1} B_k + \sum_{j=1}^{k-2} c_{j+1} B_j,\,\, c_k = 0 \end{array} \right .
$$
where, the inequalities follow from induction hypothesis and Equation~\ref{eq:crecur}. The RHS corresponds to all the admissible $c_j$s and the proof follows. \\
($\Longrightarrow$ case) This follows by contradiction and reversing the previous arguments. \proofover

\proof ({\bf Lemma~\ref{lmm:card}}) Any integer $q \in \ca_{k,c}$ has an Ostrowski decomposition:
$$
q=\sum_{j=k}^{\g_n^*} c_{j+1}(q)B_j \equiv (0,\,\ldots,\,0,c_{k+1}(q),\,c_{k+2}(q),\ldots,c_{\g_n^*+1}(q)),\,\,c_{k+1}=c
$$
where $c_{k+l}$s are arbitrary integers constrained only by property (C). Consequently it follows from Proposition~\ref{prp:order}, for any other integer, $p \in \ca_{k,c},\,\,p> q$ that,
$$
\exists \, l \geq k+2 \,\,\mbox{such that}\,\, c_l(p) > c_l(q),\,\,\& \,\,\,c_j(p) \geq c_j(q),\,\,\forall \, j \geq l
$$
This implies that,
\begin{eqnarray}\label{eq:card}
p-q &=& \sum_{j=k+1}^{\g_n^*} (c_j(p)-c_j(q))B_{j-1}=\sum_{j=k+2}^{\g_n^*} (c_j(p)-c_j(q))B_{j-1} \geq B_{l-1} - \sum_{j=k+2}^{l-1} c_j(q) B_{j-1}\\ \nom & \geq& a_{l-1} B_{l-2}-\sum_{j=k+2}^{l-1} c_j(q) B_{j-1} \geq B_{l-2}-\sum_{j=k+2}^{l-2} c_j(q) B_{j-1} \\ \nom &\geq& \ldots \geq B_{k+2} - c_{k+2}B_{k+1} \geq (a_{k+2}-c_{k+2})B_{k+1} \geq B_{k+1}
\end{eqnarray}
This means that there can only be one term belonging to $\ca_{k,c}$ for any sequential set of $B_{k+1}$ integers. Now $n$ can be written as:
$$
n=\left \lfloor{n \o B_{k+1}}\right \rfloor B_{k+1} + r,\,\, 0\leq r < B_{k+1}
$$
Therefore, the remainder, $r$, terms can contain atmost one term. This implies,
$$
\#\ca_{l,c} \leq \left \lfloor{n \o B_{l+1}}\right \rfloor + 1
$$
Now, since $\g_n^*$ is largest possible type in $Z_n$ we see that $B_{\g_n^*+1}$ must be larger than $n$. This implies that,
$$
\#\ca_{\g_n^*,c} \leq  1
$$
For all other types, $0\leq \g(m) \leq \g_n^*-1$ we have $B_\g(m)\leq n$, which implies,
$$
\#\ca_{l,c} \leq \left \lfloor{n \o B_{l+1}}\right \rfloor + 1 \leq 2\left \lfloor{n \o B_{l+1}}\right \rfloor,\,\, 0\leq \g(m) \leq \g_n^*-1
$$
Finally, in order to compute the bounds for $\#\ca_{k,c,d}$ we observe that the first two Ostrowski expansion coefficients for any two integers $p,\,q \in \ca_{k,c,d}$ are identical. This implies,
$$
p-q = \sum_{j=k+1}^{\g_n^*} (c_j(p)-c_j(q))B_{j-1}=\sum_{j=k+3}^{\g_n^*} (c_j(p)-c_j(q))B_{j-1}
$$
and the rest of proof follows as in Equation~\ref{eq:card}. \proofover
\proof ({\bf Theorem~1})
The proof of Theorem 1 follows by combining lemmas~\ref{lmm:card}~\ref{lem.lbd}.
\bqry \nom
{1\o n}\sum_{k=0}^n  {1 \o 1+n\|k\beta\|}& = &{1\o n}\sum_{k=0}^{\g_n^*} \sum_{m: \g(m)=k} {1 \o 1+n\|m\beta\|}\\ \nom &=&
{1\o n}\sum_{m\in \bigcup_{c=1}^{a_1}\ca_{0,c}}{1 \o 1+n\|m\beta\|}+ {1\o n}\sum_{k=1}^{\g_n^*-1} \sum_{m\in\bigcup_{c=1}^{a_{k+1}}\ca_{k,c}}{1 \o 1+n\|m\beta\|}\\&&+ {1 \over n} \sum_{m\in\bigcup_{c=1}^{a_{\g_n^*+1}}\ca_{\g_n^*,c}}{1 \o 1+n\|m\beta\|}
\label{e.threeterms}
\eqry
Next, we simplify each of the three terms. First we compute the contribution for the second term:
\bqry \nom
{1\o n}\sum_{k=1}^{\g_n^*-1} \sum_{m\in\bigcup_{c=1}^{a_{k+1}}\ca_{k,c}}{1 \o 1+n\|m\beta\|} &=& {1\o n}\sum_{k=1}^{\g_n^*-1} \sum_{m\in\ca_{k,1}}{1 \o 1+n\|m\beta\|} + {1\o n}\sum_{k=1}^{\g_n^*-1} \sum_{m\in\bigcup_{c=2}^{a_{k+1}}\ca_{k,c}}{1 \o 1+n\|m\beta\|} \\ \nom &&\\ \nom &\stackrel{(a)}{\leq}& {1\o n}\sum_{k=1}^{\g_n^*-1} \sum_{c=2}^{a_{k+1}} {\card \left ( \ca_{k,c}\right ) \o 1+n {c-1 \o B_{k+1}} }+{1\o n}\sum_{k=1}^{\g_n^*-1} \sum_{c=1}^{a_{k+2}-1} {\card \left ( \ca_{k,1,c}\right ) \o 1+n {a_{k+2}-c \o B_{k+2}} }\\ \nom && \\ \nom &\stackrel{(b)}{\leq}&
{1\o n}\sum_{k=1}^{\g_n^*-1} \sum_{c=2}^{a_{k+1}} {2\left \lfloor {n \o B_{k+1}} \right \rfloor  \o 1+n {c-1 \o B_{k+1}} }+{1\o n}\sum_{k=1}^{\g_n^*-1} \sum_{c=1}^{a_{k+2}-1} {2\left \lfloor {n \o B_{k+2}} \right \rfloor \o 1+n {a_{k+2}-c \o B_{k+2}} } \\ \nom && \\ \nom & \stackrel{(c)}{\leq}&
{1\o n}\sum_{k=1}^{\g_n^*-1} \sum_{c=2}^{a_{k+1}} {2 \o c-1 }+ {1\o n}\sum_{k=1}^{\g_n^*-1} \sum_{c=1}^{a_{k+2}-1} {2 \o  a_{k+2}-c } \\ \nom &&\\ \nom &\stackrel{(d)}{\leq}& {1\o n}  \sum_{k=1}^{\g_n^*-1} 2\log(a_{k+1}) + {1\o n}  \sum_{k=1}^{\g_n^*-1} 2\log(a_{k+2}) \leq {4\g_n^* \log(n) \o n} \stackrel{(e)}{\leq} {4(\log (n))^2 \o n}
\eqry
The inequality (a) follows from bounds in Lemma~\ref{lem.lbd}. We utilize the fact that $1/(1+ x + y) \leq 1/(1+x)$ for positive $x$ and $y$ to split the sum in two parts with terms belonging to $\ca_{k,c}$ and $\ca_{k,1,c}$ sets. Inequality (b) follows from Lemma~\ref{lmm:card}.  Inequality (e) follows from upper bound for $\gamma_n^*$ derived in Proposition~\ref{prp.maxtype}.

Now to simplify the first term of Equation~\ref{e.threeterms} we note that:
\begin{equation} \label{e.termzero}
{1\o 1+ n\|m\beta\|} = {1\o 1+n\min \left ({c_1-1\o 2 B_1}+{a_2-c_2 \o 2 B_2}, {a_1- c_1 \o 2B_1} +{c_2 \o 2 B_2}\right )} \leq {1\o 1+ n{a_1- c_1 \o 2B_1} +n{c_2 \o 2 B_2}}+{1\o 1+ n{c_1-1\o 2 B_1}+n{a_2-c_2 \o 2 B_2}}
\end{equation}
Bounds for the two terms on the right now follows in an identical fashion as the steps for the first term derived above.

For the last term in Equation~\ref{e.threeterms} we have two cases to consider: $\g(m) = \g_n^* \geq 1$ and $\g(m) = \g_n^* = 0$. For both cases from Lemma~\ref{lmm:card} the terms of type $\g_n^*$ are of the form $B_{\g_n^*},\,2B_{\g_n^*},\,\ldots,\,d_*B_{\g_n^*}$ of which $B_{\g_n^*}$ is the first term in $Z_n$.
Now if, $\al$, is a quadratic algebraic number and $\beta=\tau\al \in {\cal S}(n,\al)$, we know that
\begin{equation} \label{e.algebraic}
\|B_{\g_n^*}\beta\|=\|B_{\g_n^*}\tau\alpha\| \geq {C_{\alpha} \o B_{\g_n^*}\tau}
\end{equation}
Since, the last term, $cB_{\g_n^*}$ of type $\g_n^*$ has to be smaller than $n$, we have,
$$
d_* \leq \left \lfloor {n \o B_{\g_n^*}} \right \rfloor
$$
Since, $\beta=\tau\al$ for some $0<\tau \leq n$, we have,
\bqry \nom
\sum_{m\in\bigcup_c\ca_{\g_n^*,c}}{1 \o 1+n\| m\tau \al \|} &\leq & {1\o n}\sum_{j=1}^{d_*} {1 \o 1+n j{C_\al \o B_{\g_n^*}\tau}} \leq {1\o n}\sum_{j=1}^{d_*} {1 \o 1+ j {C_\al \o B_{\g_n^*}}}= {1\o n}\sum_{j=1}^{d_*} {B_{\g_n^*} \o B_{\g_n^*}+ C_\al j } \\ \nom && \\ \nom &\leq & {1\o n}\min \left(d_*,{B_{\g_n^*} \o C_\al}\log \left (1+ d_*) \right ) \right )\\ &\leq & {1\o n}\min \left(\left \lfloor {n \o B_{\g_n^*}}\right \rfloor ,{B_{\g_n^*} \o C_\al}\log \left (1+{n \o B_{\g_n^*}} \right ) \right ) \leq  {\log^{0.5}(n)\o C_\al^{0.5} \sqrt{n}}
\label{e.termthree}
\eqry

Now for the case when $\g(m)=\g_n^* = 0$ we invoke Corollary~\ref{cor.lbd} and follow along the lines of the proof for $\g(m)=0$ of Equation~\ref{e.termzero} by first splitting it into two terms, i.e.,
\begin{align*}
{1\o 1+ n\|m\beta\|} &=& {1\o 1+n\min \left ({(c_1-1)|D_0|}+{(a_2-c_2)|D_1|}, {(a_1- c_1)|D_0|} +{c_2|D_1|}\right )}\\& \leq & {1\o 1+ n{(a_1- c_1)|D_0|}}+{1\o 1+ n{(c_1-1)|D_0|}}
\end{align*}
We now follow along the lines of Equation~\ref{e.termthree} by first substituting for $D_0$ given by Equation~\ref{e.algebraic} and following along the lines of Equation~\ref{e.termthree}.
To complete the proof we need to ensure that the PDACF property continues to hold for $0< k, \tau \leq p=\lambda n$, with $\lambda > 1$. To do this we again analyze Equation~\ref{e.threeterms}. The first two terms depends only on the cardinality of the set ${\cal A}_{k,c}$. Following along the lines of Lemma~\ref{lmm:card} we see that,
$$
\#\ca_{l,c} \leq \left \{ \begin{array}{cc} 2 \lambda \left \lfloor{n \o B_{l+1}}\right \rfloor & l < \g_n^* \\ 1 & l=\g_n^* \end{array} \right .
$$
For the third term we observe that,
$$
\sum_{m\in\bigcup_c\ca_{\g_{\lambda n}^*,c}}{1 \o 1+n\| m\tau \al \|} \leq {\log^{0.5}(n)\o (\lambda C_\al)^{0.5} \sqrt{n}}
$$
\proofover

\section{Matrices with RIP Property} \label{sec:mrip}
In this section we quantify the RIP property for matrices described in Section~\ref{sec:prob}. Consider the Toeplitz constructions in Equation~\ref{eqn:toepmat}. It is clear from the sequence designs in the previous section together with Theorem~\ref{thm:c6chirp3} that the RIP property (see Definition~\ref{dfn.rip}) of order $(\lambda n, n, n^{1/4}/\lambda^{0.5})$ is satisfied. However, this bound is quite loose and we establish substantial improvement over this bound.
We have the following result:
\begin{thm}
Consider the Toeplitz construction of Equation~\ref{eqn:toepmat} with the elements generated by the HOC with $\alpha$ equal to the golden ratio. It follows that for sufficiently large $n$, the RIP property of order $(\lambda n, n, n^{3/8}/(\sqrt{\lambda} \log(n)))$ is satisfied for this matrix construction.
\end{thm}
The idea is that to establish RIP property of order $k$, by Theorem~\ref{thm:rip} we only need to show that for all subsets, $I_k \subset \{1,\,2,\,\ldots,\,\lambda n\}$ of size $k$,
\beq \label{e.ripmat}
\sum_{\tau \in I_k} |r_u^n(\tau)| \leq   \sum_{\tau \in I_k } \left ( {1 \over n} \sum_{m=1}^{\lambda n} {1  \over 1 + n \|m \tau \al \|} \right )^{1/2} < 1
\eeq
Consequently an autocorrelation decay of order $n^{-3/8}$ is only required on ``average''. Now there are contributions of three terms as seen from Equation~\ref{e.threeterms} and the first two terms already decay at this rate. Therefore, we are left to analyze the third term, namely, terms that belong to the largest type, $\gamma_n^*$ for each $\tau$. These are the only terms that contribute towards the slow decay. Therefore, we are left to establish that the number of terms with slow decay are relatively small, which we present next.


Next we decompose the RHS into three terms as in Equation~\ref{e.threeterms}. We note that the first two terms are of ${\cal O}(n^{-1})$. Therefore we can write:
\bqry \nom
{1 \over n^2} \sum_{\tau \in {\cal I}_k } \left ( \sum_{m=1}^{\lambda n} {1  \over 1 + n \|m \tau \al \|} \right )^{1/2} &\leq& {1 \over n^2} \sum_{\tau \in {\cal I}_k } \left ( {\lambda \o n} + {\lambda \o n} + \sum_{m \in \cup_c \ca_{\gamma_n^*,c}} {1  \over 1 + n \|m \tau \al \|} \right )^{1/2} \\ \nom & \leq &  \sum_{\tau \in {\cal I}_k } \sqrt{2} \max \left ( \sqrt{{2 \lambda \o n}},  \left ({1\o n} \sum_{m \in \cup_c \ca_{\gamma_n^*,c}} {1  \over 1 + n \|m \tau \al \|} \right )^{1/2}  \right ) \\ \nom
&\leq& \sqrt{2} \max \left (|{\cal I}_k|\sqrt{{2 \lambda \o n}}, \sum_{\tau \in {\cal I}_k} \left ( {1\o n} \sum_{j=1}^{\lfloor {\lambda n \o B_{\gamma_n^*}} \rfloor } {1  \over 1 + n j \|B_{\gamma_n^*} \tau \al \|} \right )^{1/2} \right )
 \\ \nom
\eqry
First note that by hypothesis we have,
$|{\cal I}_k| \leq {\sqrt{n} \o \sqrt{\lambda} \log(n)}$ and hence the first expression in the maximum is smaller than one.
We are now left to compute the contribution from terms of the largest type. Let $B_{\gamma_n^*}$ be the first term of the largest type. Note that $B_{\gamma_n^*}$ depends on $\tau$ but we suppress this dependence to simplify notation.
Let
$$
\phi(\tau) = \sum_{j=1}^{\lfloor{\lambda n \over B_{\gamma_n^*}\rfloor } }{1 \over 1 + nj \|B_{\gamma_n^*} \tau \al\|},\,\tau = 1, \,2,\,\ldots,\,n
$$

%


We have the following lemma that provides decay bounds for $\phi(\tau)$:
\begin{lemma} \label{lmm.decbds}
Suppose $\tau$ is such that $B_{\gamma_n^*} \tau \geq n^{\eta_1}$ and $\|B_{\gamma_n^*} \tau \al \| = n^{-\eta_2}$ for some $\eta_1,\,\eta_2 \in [1,\,2]$ then,
\beq \label{e.minupbd}
\phi(\tau) \leq n^{\min\{1-\eta_1, \eta_2-2\}} \log(n)
\eeq
for $1 \leq \eta_2 \leq \eta_1$ and zero if $\eta_2 > \eta_1$. Furthermore, if $\eta_2 < 1$ then,
$$
\phi(\tau) \leq {\lambda \o C_\al n}
$$
\end{lemma}
\proof
The last part is immediate and follows from:
$$
\phi(\tau) = {1\o n} \sum_{j=1}^{\lfloor {\lambda n \o B_{\gamma_n^*}} \rfloor } {1  \over 1 + n j \|B_{\gamma_n^*} \tau \al \|} \leq {\lambda \log(n)\o C_\al n}
$$
To establish the rest of the statements we note the fact that $\al$ is a quadratic irrational. This imposes the fact that if $B_{\gamma_n^*}\tau = n^{\eta_1}$ then $\|B_{\gamma_n^*}\tau \al \| \geq C_{\al} n^{-\eta_1}$. Next note that $\lfloor {\lambda n \o B_{\gamma_n^*}} \rfloor  \leq n^{2-\eta_1}$ and $n \|B_{\gamma_n^*} \tau \al \| \geq n^{\eta_2-1}$. Consequently, when $\eta_1 - 1 > 2-\eta_2$ we use the fact that
$${1  \over 1 + j n \|B_{\gamma_n^*} \tau \al \|} \leq 1$$
On the other hand when $\eta_2 - 1 < 2-\eta_1$ we have
$${1  \over 1 + j n \|B_{\gamma_n^*} \tau \al \|} \leq {1 \o j n^{1-\eta_2}}$$
\proofover
The above lemma immediately implies that if $\tau$ is such that $\eta_2 < 1$ and  ${\cal I}_k \leq {C_{\al} \o \sqrt{2} \lambda} \sqrt{n}$ the autocorrelation for this shift terms, $\tau$, will be negligible.

Note that the minimum in the RHS of Equation~\ref{e.minupbd} is achieved for $\eta_1=\eta_2=1.5$. For this value we get $\phi(\tau) \leq \log(n)/\sqrt{n}$. To understand the contributions for different $\tau$ we partition it into different decay rate regions. To this end let, $0.5=\theta_1 < \theta_2 < \ldots < \theta_L=1$ be a partition of the interval $[0.5,\,1]$ with $\delta = \theta_{j+1}-\theta_j$ and $L$ a large fixed integer independent of $n$. Let
$$
{\cal S}_j = \{ \tau \mid \phi(\tau) \in [n^{-\theta_j},\, n^{-\theta_{j-1}}]\},\,\,\tau=1,\,2,\,\ldots,\,n
$$
The cardinality of these sets is given in the following lemma:
\begin{lemma}
Let $\al$ be the golden ratio, i.e., $\al = (1+\sqrt{5})/2$ then for sufficiently large $n$,
$$
|{\cal S}_j| \leq n^{1.14(2\theta_j-1)}
$$
\end{lemma}
\proof
  Let \beq \label{e.hyp0}
  \phi(\tau) \in [n^{-\theta_{j+1}},\,\,n^{-\theta_j}]\eeq It follows from Lemma~\ref{lmm.decbds} that,
  $$
  \eta_1 \leq 1+\theta_{j+1},\,\,\eta_2 \geq 2-\theta_j
  $$
Noting that $n^{\eta_1} = B_{\gamma_n^* \tau}$ and $n^{-\eta_2} = \|B_{\gamma_n^*}\tau \al\|$ and the fact that $\al$ is a quadratic irrational implies that  $\|B_{\gamma_n^*}\tau \al\| \geq C_{\al} n^{-\eta_1}$.  By substituting for $\eta_1$ and $\eta_2$ we get:
  \beq \label{e.cl1}
  \|B_{\gamma_n^*} \tau \al \|\in [n^{-(1+\theta_{j+1})}, n^{-(2-\theta_j)}]
  \eeq
In addition we have,
$$
B_{\gamma_n^* \tau} = n^{\eta_1} \leq n^{1+\theta_{j+1}}
$$
Motivated by Equation~\ref{e.cl1} we denote by
$$
{\cal C}_j = \{ m \in Z^+ \mid \|m \al \| \in [n^{-(1+\theta_{j+1})}, n^{-(2-\theta_j)}],\,\,m \leq n^{1+\theta_{j+1}} \}
$$

By construction note that for $\tau \in {\cal S}_j$ and associated largest type $B_{\gamma_n^*}$ the product $B_{\gamma_n^*} \tau \in {\cal C}_j$. Thus our proof now relies on estimating the cardinality of ${\cal C}_j$ and then computing the number of divisors $\tau$ for each element of ${\cal C}_j$. In other words,
\beq \label{e.card}
|{\cal S}_j| \leq |{\cal C}_j| \max_m \{\mbox{ \# divisors for $m \in {\cal C}_j$} \}
\eeq

Let $F_k,\,\,k=0,\,1,\,\ldots$ be the convergents of the golden ratio. It is well known that the convergents satisfy a linear second order recursion
\beq \label{e.recur}
F_{k+1} = F_k + F_{k-1},\,\, F_0 = F_1 =1
\eeq
The continued fraction expansion of the golden ratio is:
$$
\al=[1;1,\, 1,\,1,\,\ldots]
$$
A closed form expression for $F_k$ is given by
\beq \label{e.closedform}
F_k = {\al^k - 1/\al^k \over \sqrt{5}}
\eeq
where $\al = (1+\sqrt{5})/2$. From Lemma~\ref{lem.lbd} it follows that if $m \in {\cal C}_j$  then the type of $m$ in the Ostrowski representation (i.e. the smallest element) must satisfy:
\beq \label{e.smalltype}
F_{\gamma(m)} \in [[n^{(1+\theta_{j+1})}, n^{(2-\theta_j)}]]
\eeq
Now the largest element in ${\cal C}_j$ is clearly $m=n^{1+\theta_{j}}$. Consequently, the number of different convergents, $s$ in the Ostrowski representation of the set ${\cal C}_j$ can be obtained by employing Equation~\ref{e.closedform} and Equation~\ref{e.smalltype}:
$$
s \leq {(2 \theta_j+\delta -1)\log(n)\over \log(\al)} + 1
$$
A crude upper bound for the cardinality is clearly $2^s$ since the set of all expansions can be associated with $s$ binary digit expansions. However, from Equations~\ref{e.recur},~\ref{eq:ostr} we see that no two consecutive convergents can be present in any expansion. Therefore, the cardinality of the set can be refined and given by the following expression:
$$
|{\cal C}_j| \leq \sum_{k=0}^{\lfloor s/2 \rfloor} {s-2k \choose k} \leq \max_{\omega \in [0,\,1]} 2^{\lfloor s/2 \rfloor (\omega + H_2(\omega))}
$$
where $H_2(\cdot)$ is the binary entropy function. The RHS is obtained through Stirling's approximation. Further simplification yields:
$$
\sum_{j=0}^{\lfloor s/2 \rfloor} {s-2j \choose j} \leq 2^{0.55 s} = 2^{0.55 (2\theta_j + \delta -1) \log(n)/\log(\al)} \approx n^{0.78(2\theta_j+\delta-1) }
$$
Finally, to compute the cardinality of ${\cal S}_j$ we see from Equation~\ref{e.card} we need to determine the number of divisors for each $m \in {\cal C}_j$.
This is a classical result going back to Ramanujan~\cite{ramanujan}:
\begin{lmm} \label{lmm:ramanujan}
For an arbitrary $\eps>0$, there exists a positive integer $m_0$ such that the number of divisors is smaller than $2^{(1+\eps)\log(m)/\log\log(m)}$ for all $m \geq m_0$.
\end{lmm}
Consequently, the cardinality of ${\cal S}_j$ for sufficiently large $n$ is given by:
\beq \label{e.numoftypes}
|{\cal S}_j| \leq 2^{(1+\eps)\log(n)/\log\log(n)} n^{0.78 (2\theta_j+\delta-1)} \leq n^{0.78 (2\theta_j+\delta-1) + \upsilon}
\eeq
where $\upsilon$ is an arbitrary small number for sufficiently large $n$.
\proofover

Now to complete the proof of the main theorem we determine the set of partitions for which the contributions for $\tau \in \{1,\,2,\,\ldots,\,n\}$ approaches zero, i.e.
$$
\min_{k} \sum_{j \geq k} \sum_{\tau \in {\cal S}_j} \phi^{1/2}(\tau) \leq \min_k \sum_{j\geq k} |{\cal S}_j| n^{-\theta_j} \longrightarrow 0
$$
Upon direct substitution it follows that this is ensured whenever, $0.78(2\theta_j-1)\leq \theta_j/2$ (ignoring the small $\delta$ and $\upsilon$ terms). Upon evaluation we infer that this holds for all $\theta_j \leq 3/4$. We now let $j_{\min}$ be the minimum $j$ such that $\theta_j \leq 3/4$.
Consequently, we are left with
$$
\sum_{\tau \in {\cal I}_k} \phi^{1/2}(\tau) \leq \sum_{j=1}^{j_{\min}} \sum_{\tau \in {\cal I}_k\cap {\cal S}_j} \phi^{1/2}(\tau) + \sum_{j \geq j_{\min}} \sum_{\tau \in {\cal S}_j} \phi^{1/2}(\tau) \leq |{\cal I}_k| j_{\min} \max_{\tau \in {\cal S}_j,\,j \leq j_{\min}}  \phi^{1/2}(\tau) + {\cal O}(\eps)
$$
where $\eps>0$ can be chosen to be an arbitrary small positive number for sufficiently large $n$. Note that $\theta_j \leq 3/8$ implies $\phi(\tau) \leq n^{-3/8}$. Therefore, if $$|{\cal I}_k| \leq {n^{3/8} \over \log(n)}$$ the RHS can be made arbitrarily small as well and the result follows.
\proofover

We observe that the golden ratio, $\al$, provides a moderate improvement in RIP property. We point to Equation~\ref{e.numoftypes} as one of the principle reasons. This Equation shows that number of large type of order $m=n^{\kappa}$ betweeen two (large) numbers $m_1 = n^{\kappa_1}$ and $m_2=n^{\kappa_2}$ scales polynomially with $n$. This can be attributed to the logarithmic scaling ($log(n)$) of the number of convergents between $m_1$ and $m_2$. Consequently, a question that arises is whether one could control this scaling by choosing a different quadratic irrational number. One possibility is to choose recursions (see Equation~\ref{eq:crecur}) that lead to fewer convergents. This can be accomplished by choosing periodic continued fraction expansions of the form: $\al = [a_0; a_1]$ where $a_1$ is large. However, fewer convergents does not result in decreasing the number of integers of large type since now we can admit $B_{\gamma_n}^*\tau \al,\,2B_{\gamma_n}^*\tau \al,\,\ldots,\,a_1B_{\gamma_n}^*\tau \al$ as alternatives. Therefore, we conjecture that this is nearly the best result one could hope for with quadratic irrational numbers.

\section{Discussion}
\begin{figure}
\begin{center}
\resizebox{!}{3in}{\includegraphics{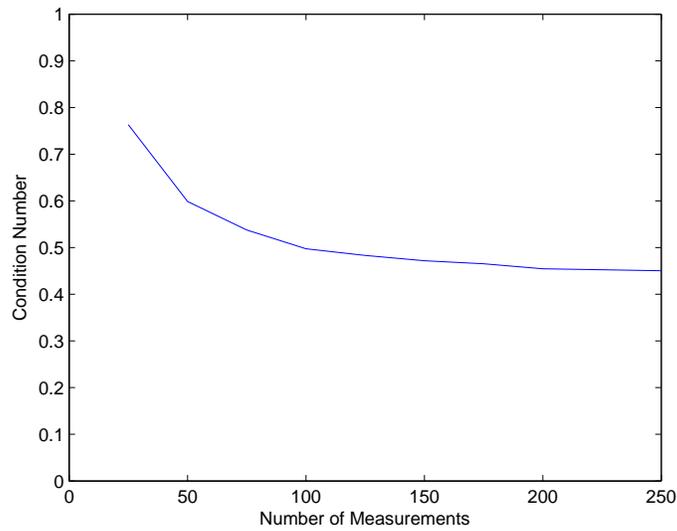}}\\
\caption{The condition number (ratio of maximum to minimum singular values) is plotted as a function of number of measurements. 5000 Monte Carlo simulations were performed with the number of measurements ranging from $50$ to $500$ for Toeplitz construction of Equation~\ref{eqn:toepmat}. The number of variables $p$ was scaled as twice the number of measurements. The sparsity level was scaled as $20 \%$ of the number of measurements.  }
\label{f.ripexp}
\end{center}
\end{figure}

We first point out that the RIP order derived in the previous section appears to be small relative to what can be obtained with unstructured constructions~\cite{calderb}. We believe that this looseness is inherently a consequence of the bounding technique utilized here. To further confirm our belief we performed a Monte Carlo simulation to determine the condition number (ratio of the maximum to the minimum singular value) for various choices of sparsity levels for Toeplitz matrix of Equation~\ref{eqn:toepmat}. In particular we chose the $(n,p,q)$ parameters as follows: We let $n$ denote the number of measurements; we set the number of variables $p = 2n$ and the sparsity level $q = n/5$. The results are described in Figure~\ref{f.ripexp}.

We next briefly discuss various implementational and practical issues here.

First, note that we have described properties for complex-valued signals. Nevertheless it is straightforward to check that identical properties hold for real and imaginary parts of the higher-order-chirp(HOC) signals.
A second claim that can be immediately verified is one of doppler resilience. This is because the autocorrelation properties of HOCs are not affected by constant frequency shifts.
A third aspect is that our construction of Toeplitz matrices could be employed in sequential processing. This is because each time a new row is added in Equation~\ref{eqn:toepmat} or Equation~\ref{eqn:fulltoep} the new matrix formed by concatenation of the previous matrix with the new row still preserves the RIP property. Finally these sequences and matrices can be generated with relatively little memory. To see this note that since the autocorrelations decay at a polynomial rate, $n^{-\gamma}$, for a toeplitz matrix of $n$ rows, we only need to ensure that our approximation does not have larger error than this magnitude. Again continued fraction expansion can be employed for this task. For instance, consider the HOC
$$
u_t=\exp(-j2 \pi \al t^3) = \exp(-j2 \pi \|\al t^3\|)
$$
For the golden ratio, $\al=(1+\sqrt{5})/2$ we can compute the above expression to any order of approximation conveniently. This is because the convergents of $\al$ are given by the Fibonacci series, $F_k$, which can be computed through simple recursion (see Equation~\ref{e.recur}). Furthermore, to compute a rational approximation to $q\al$ for any $q \in Z$ we can proceed as follows. Let $F_q$ be any term in the Fibonacci series for which $q$ is a factor (note that there are infinitely many such terms for every whole number $q$ \cite{vorob}). Then
$$
\min_p |\al - {p \o F_q}| \leq {1 \over \sqrt{5} F_q^2} \implies  \min_p |q\al - {qp \o F_q}| \leq {q \over \sqrt{5} F_q^2} \implies q\al \in \left [{qp \o F_q} \pm {q \over \sqrt{5} F_q^2}\right ]
$$
Since $F_q$ can be chosen to be sufficiently large we can get arbitrarily good approximations through employing Fibonacci series.

\bibliography{references,onr_bib}
\end{document}